\pdfoutput=1
\documentclass[letter,american,citeautoscript,floatfix,pdftex,superscriptaddress,twocolumn,twoside,%
aps,prb,%
reprint
]{revtex4-1}
\usepackage{amsfonts,amsmath,amssymb}
\usepackage{commath}%
\usepackage[T1]{fontenc}
\usepackage{graphicx}
\usepackage[utf8]{inputenc}
\usepackage{microtype}
\usepackage{xspace}
\usepackage{hyperref}
\usepackage[notodos]{CMImacros}

\newlength{\figwidth}
\newcommand{\cfeldesy}{\affiliation{Center for Free-Electron Laser Science, Deutsches
      Elektronen-Synchrotron DESY, Notkestrasse 85, 22607 Hamburg, Germany}}%
\newcommand{\uhhcui}{\affiliation{The Hamburg Center for Ultrafast Imaging, Universit\"at Hamburg,
      Luruper Chaussee 149, 22761 Hamburg, Germany}}%
\newcommand{\uhhphys}{\affiliation{Department of Physics, Universit\"at Hamburg, Luruper Chaussee
      149, 22761 Hamburg, Germany}}%

\begin{document}
\title{Development and characterization of a laser-induced acoustic desorption source}
\author{Zhipeng Huang}\cfeldesy\uhhphys%
\author{Tim Ossenbrüggen}\cfeldesy%
\author{Igor Rubinsky}\cfeldesy\uhhcui%
\author{Matthias Schust}\cfeldesy%
\author{Daniel A. Horke}\cfeldesy\uhhcui%
\author{\mbox{Jochen Küpper}}%
\email[]{jochen.kuepper@cfel.de}%
\homepage{https://www.controlled-molecule-imaging.org}%
\cfeldesy\uhhphys\uhhcui%
\date{\today}%
\begin{abstract}\noindent
   A laser-induced acoustic desorption source, developed for use at central facilities, such as
   free-electron lasers, is presented. It features prolonged measurement times and a fixed
   interaction point. A novel sample deposition method using aerosol spraying provides a uniform
   sample coverage and hence stable signal intensity. Utilizing strong-field ionization as a
   universal detection scheme, the produced molecular plume is characterized in terms of number
   density, spatial extend, fragmentation, temporal distribution, translational velocity, and
   translational temperature. The effect of desorption laser intensity on these plume properties is
   evaluated. While translational velocity is invariant for different desorption laser intensities,
   pointing to a non-thermal desorption mechanism, the translational temperature increases
   significantly and higher fragmentation is observed with increased desorption laser fluence.
\end{abstract}
\maketitle

\section{Introduction}
\label{sec:introduction}
Recent years have seen the development of several techniques to control isolated neutral molecules
in the gas-phase. Molecular beams of polar molecules can be dispersed with strong inhomogeneous
electric fields, producing pure samples of individual conformers, cluster stoichiometries or even
single quantum-states~\cite{Filsinger:PRL100:133003, Filsinger:ACIE48:6900, Trippel:PRA86:033202,
   Nielsen:PCCP13:18971, Horke:ACIE53:11965, Meerakker:CR112:4828, Chang:IRPC34:557}. We can,
furthermore, control the alignment and orientation of complex gas-phase molecules in
space~\cite{Holmegaard:PRL102:023001, Filsinger:JCP131:064309, Trippel:MP111:1738,
   Stapelfeldt:RMP75:543}, allowing one to extract molecular-frame information, such as nuclear or
electronic structures, from these samples~\cite{Bisgaard:Science323:1464, Holmegaard:NatPhys6:428}.
In combination with the technological developments in free-electron laser (FEL) ultrafast x-ray
sources, now providing millijoule-level pulses of hard x-rays with sub-100~fs pulse durations, these
control techniques open up the potential to image isolated biomolecules and particles with
femtosecond temporal and picometer spatial resolution~\cite{Seibert:Nature470:78,
   Neutze:Nature406:752, Boll:PRA88:061402, Kuepper:PRL112:083002}.

The realization of these experiments crucially depends on a high-density source of intact molecules
in the gas-phase, ready for further manipulation and experiments. While for many small stable
compounds this is easily achieved using thermal vaporization and seeding into a molecular beam, this
approach is not feasible for thermally labile or non-volatile species -- such as most larger
biochemically relevant molecules, and biological species in general. Therefore, these samples
require the development of gentle vaporization techniques, that still produce a pure and
high-density sample of molecules in the gas-phase. Furthermore, technical requirements for
central-facility experiments, such as a well-defined and fixed interaction point and capabilities
for long uninterrupted measurement times, need to be fulfilled.

One approach to achieve relatively dense ensembles of labile neutral molecules is laser-induced
acoustic desorption (LIAD), which has been introduced over 30 years
ago~\cite{Lindner:AnalChem57:895}, but received relatively little attention since. What sets LIAD
apart from other laser-based vaporization techniques, such as laser
desorption~\cite{DeVries:ARPC58:585}, is that it avoids any direct interaction between the
desorption laser and the molecular sample, making this technique applicable to light-sensitive and
labile compounds. The basic principle of LIAD is that samples get deposited on one side of an opaque
substrate -- often a thin metal foil -- while the other side of this substrate gets irradiated with
a laser pulse. This laser pulse induces acoustic and thermal waves within the
substrate, which travel through the material and lead to desorption of molecules on the front side.
The physical mechanism behind this desorption process is currently very poorly understood, \ie, even
the nature of the desorption process (thermal, acoustic, stress-induced) is not clearly established
and, furthermore, it is highly dependent on the employed substrate and sample preparation
method~\cite{Zinovev:AnalChem79:8232}.

Nonetheless, the LIAD technique has been used in a number of mass spectrometry
studies~\cite{Golovlev:ijmsip169:69, Peng:ACIE45:1423, Nyadong:AnalChem84:7131}. Notably, the
Kenttämaa group coupled LIAD to a Fourier transform ion cyclotron mass
spectrometer~\cite{Perez:IJMS198:173, Shea:AnalChem79:2688, Shea:AnalChem78:6133} and a quadrupole
linear time-of-flight mass spectrometer~\cite{Habicht:AnalChem82:608, Gao:JASMS22:531}. They used
this source to study peptides and large organic compounds up to mass $\ordsim500$~u. Recently, the
LIAD methodology has also been applied to study the dynamics of intact aminoacids on the femtosecond
and attosecond timescale using ion-yield and photoelectron spectroscopy~\cite{Calvert:PCCP14:6289,
   Belshaw:JPCL3:3751, Calegari:Science346:336}. In a seminal paper in 2006, Peng \etal showed the
applicability of LIAD to significantly larger systems and particles, successfully desorbing viruses,
bacteria and cells and storing them in a quadrupole ion trap for precise mass
measurements~\cite{Peng:ACIE45:1423, Zhang:AnalChem88:5958}. The Campbell group furthermore
established a closely related technique, termed ``laser-induced forward transfer'' for the gentle
vaporization of large nanoparticles~\cite{Bulgakov:JOSAB31:C15, Goodfriend:APPA122:154}.

Here, we present our new LIAD-source setup, designed for use in central facilities. It allows for
prolonged measurement times through automatic sample replenishment, whilst keeping the interaction
point fixed. This is realized through the use of a long metal tape as the LIAD substrate, which is
constantly forwarded -- akin to an old-fashioned cassette tape -- to provide fresh sample. A
reproducible layer of molecules is prepared on this foil by spraying aerosolized samples onto the
band. This technique yields a stable and reproducible signal for many hours of measurement time. As
a test system we use the amino acid phenylalanine and characterize the produced molecular plume
using strong-field ionization, evaluating the number density, spatial extend and temporal
distribution. By convoluting the initial plume temporal distribution with a Maxwell-Boltzmann
velocity distribution, the forward velocity and the translational temperature in the moving frame
were derived. While the velocity does not increase with desorption laser intensity, the
translational temperature does increase and, furthermore, we observe enhanced fragmentation. These
observations are consistent with a desorption model based on surface stress between the foil band
and islands of deposited molecules, which was previously proposed~\cite{Zinovev:AnalChem79:8232}.

\section{Experimental method}
\label{sec:Experimental method}
\begin{figure}
   \centering
   \includegraphics{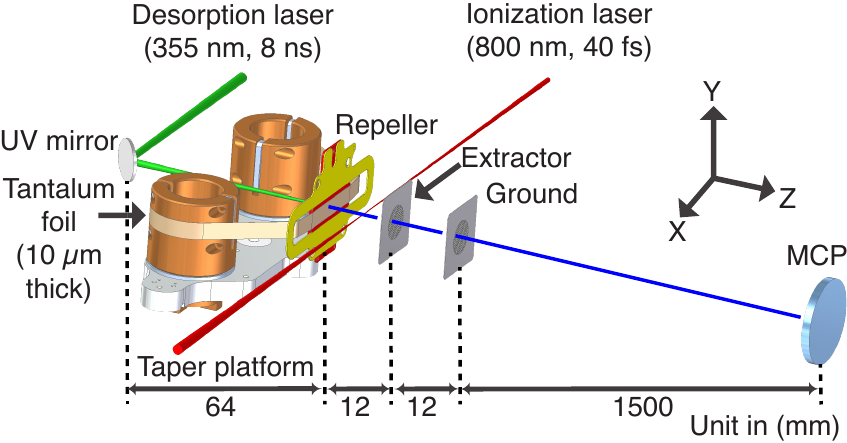}
   \caption{LIAD setup with sample delivery based on a rotating tape drive. A taper platform holds a
      long metal tape with sample applied on the front surface. A UV desorption laser irradiates the
      foil from the back, desorbing molecules. These are then ionized by a femtosecond laser beam
      and detected using a time-of-flight mass spectrometer. See supplementary information for
      further details.}
   \label{fig:setup}
\end{figure}
A schematic of our new LIAD setup is shown in \autoref{fig:setup}; further details regarding the
setup and sample preparation are given in the supplementary information. Briefly, sample is
deposited on the front side of a tantalum foil band of 10~\um thickness and 10~mm width, while the
backside gets irradiated with a pulsed desorption laser. We use tantalum as a substrate
   due to its very high melting point of $3290$~K and hence its ability to withstand higher
   desorption laser intensities. During data collection the foil band is constantly moved across
the desorption laser spot to provide fresh sample, as further discussed below. In order to create a
stable coverage of sample on the foil, we aerosolized samples using a gas-dynamic virtual nozzle
(GDVN)~\cite{DePonte:JPD41:195505, Beyerlein:RSI86:125104} to create and deposit an aerosol on the
foil, where it sticks and rapidly dries out. Full details of the sample preparation and
   deposition process, including details regarding sample concentration, spray rate, speed of the
   foil band, and an estimate of total deposited material are given in the supplementary
   information.

Molecules are desorbed using $\ordsim8$~ns duration laser pulses at 355~nm, focused to a 300~\um
(FWHM) spot on the foil. Desorbed molecules are strong-field ionized by 40~fs pulses from a
Ti:Sapphire laser, with typical field strengths of $4\times10^{13}$~W/cm$^{2}$.
Produced cations are detected by a conventional linear time-of-flight mass spectrometer
(TOF-MS), with a typical mass resolution $m/\Delta{m}>1000$.

\section{Results and Discussion}
\label{sec:results}
\subsection{Characterizing LIAD by strong-field ionization}
\label{sec:SFI}
We characterize the desorbed molecular plume using strong-field ionization (SFI) from a focused
femtosecond Ti:Sapphire laser as a universal probe~\cite{Calvert:PCCP14:6289,
Teschmit:JCP147:144204}.
\begin{figure}
   \centering
   \includegraphics[width=\linewidth]{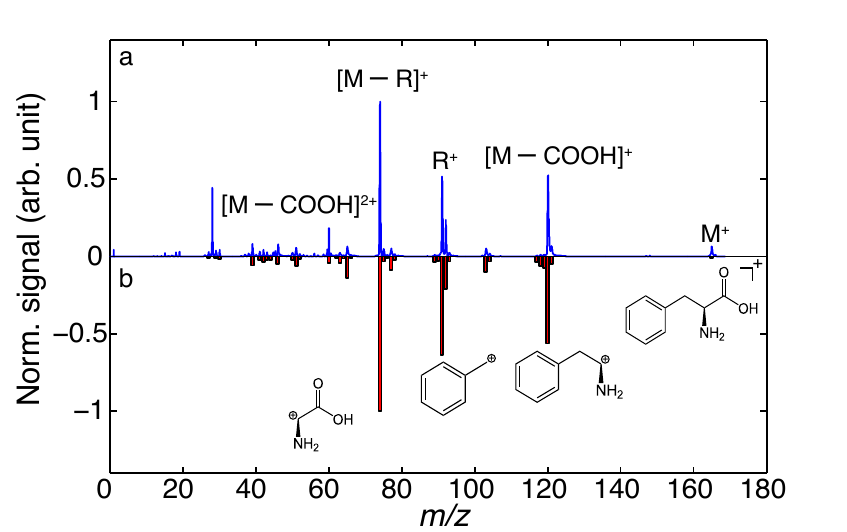}
   \caption{Mass spectrum of phenylalanine; (a) recorded using LIAD and SFI from a femtosecond laser
      beam and (b) reference spectrum for electron impact ionization~\cite{NIST:webbook:2017}. The
      intensity in both spectra is normalized to the dominant mass peak at 74~u.}
   \label{fig:tof}
\end{figure}
The observed TOF-MS of phenylalanine (PA) is shown in \autoref{fig:tof} and compared to a literature
spectrum obtained using electron-impact ionization (EI)~\cite{NIST:webbook:2017}. Both spectra are
normalized to the most abundant fragment ion at mass 74~u, corresponding to loss of
   a benzyl-radical fragment. It is evident that both ionization schemes strongly induce
fragmentation, however, we note that using SFI a significant contribution from intact PA is observed
at 165~u; this could even be enhanced using shorter duration laser
   pulses~\cite{Calvert:PCCP14:6289}. We observe no evidence for the production of larger clusters
of PA, and hence attribute this channel to desorption of intact PA monomers. Furthermore,
we observe an additional fragmentation peak at 28~u in the SFI data, corresponding to
   CNH$_2$, \eg, C-NH$_2^+$ or HC=NH$^+$, fragment ions, which is absent in the EI mass spectrum.
These spectra clearly demonstrate the production of intact PA following desorption from the foil
band. We do not observe the emission of any tantalum atoms or clusters, which would easily be
ionized by the SFI probe, since the ionization potential of tantalum is lower than of PA. This
indicates that the desorption laser does not penetrate through the foil band nor ablate metal from
the foil by other means.

\begin{figure}
   \centering
   \includegraphics{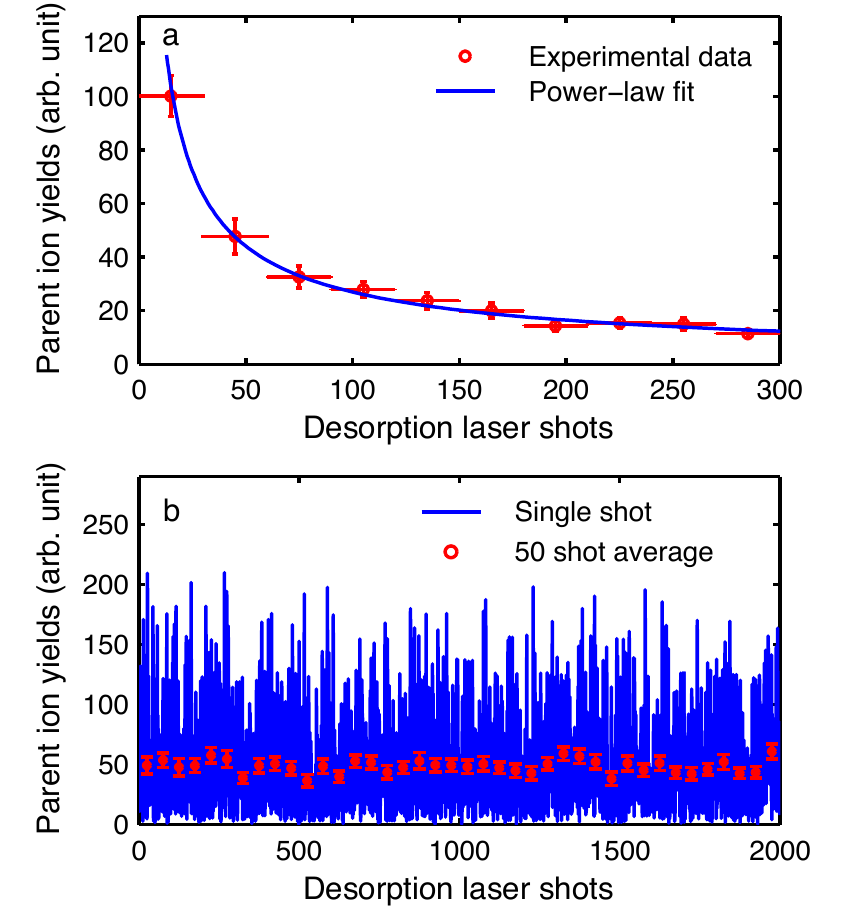}
   \caption{(a) Parent ion yield as a function of desorption laser shot without sample
      replenishment. Data have been averaged over 30~shot-wide intervals (horizontal bars); the
      solid line corresponds to a power-law fit. (b) Parent ion signal as a function of desorption
      laser shot while moving the foil band at 50~\um/s. The blue line corresponds to single-shot
      measurements, red markers correspond to averaged data for 50 shots, showing a standard
      deviation below 10\%.}
   \label{fig:stability}
\end{figure}
To assess the depletion of sample from the foil and determine the required moving speed for sample
replenishment, we measured the parent ion yield as a function of the number of desorption laser
shots onto the same spot. The resulting abundances are shown in \subautoref{fig:stability}{a},
where the solid line represent a power-law fit of the form $y = A\times{}x^n$, with an exponent of n
= $-0.68\pm0.03$. We observe a rapid decay of signal, reaching around 10\% after 330 desorption
laser shots. Similar power-law behavior has previously been observed and rationalized with the
existence of several isolated desorption centers on the foil~\cite{Zinovev:AnalChem79:8232}. This is
consistent with our observation of many large crystalline islands, see supplementary information,
many of which fall within the desorption laser spot size.

During further data collection the foil band is continuously moved at 50~\um/s, corresponding to a
movement to a new sample spot every $\ordsim120$ desorption laser shots. The corresponding
shot-to-shot signal stability for the moving foil band is shown in \subautoref{fig:stability}{b}.
The signal exhibits large fluctuations with a single shot standard deviation of 70\% of the mean
value. No long-term drift of the overall signal levels is observed. Averaging over 50 desorption
laser shots reduces the standard deviation to below 10\%, as indicated by the red markers and error
bars in \autoref{fig:stability}. Further data points in this manuscript are typically averaged over
1200 desorption laser shots, resulting in a standard deviation of $\ordsim2.5$\%.

\subsection{Molecular plume properties}
\label{sec:plume}
\begin{figure}
   \centering
   \includegraphics{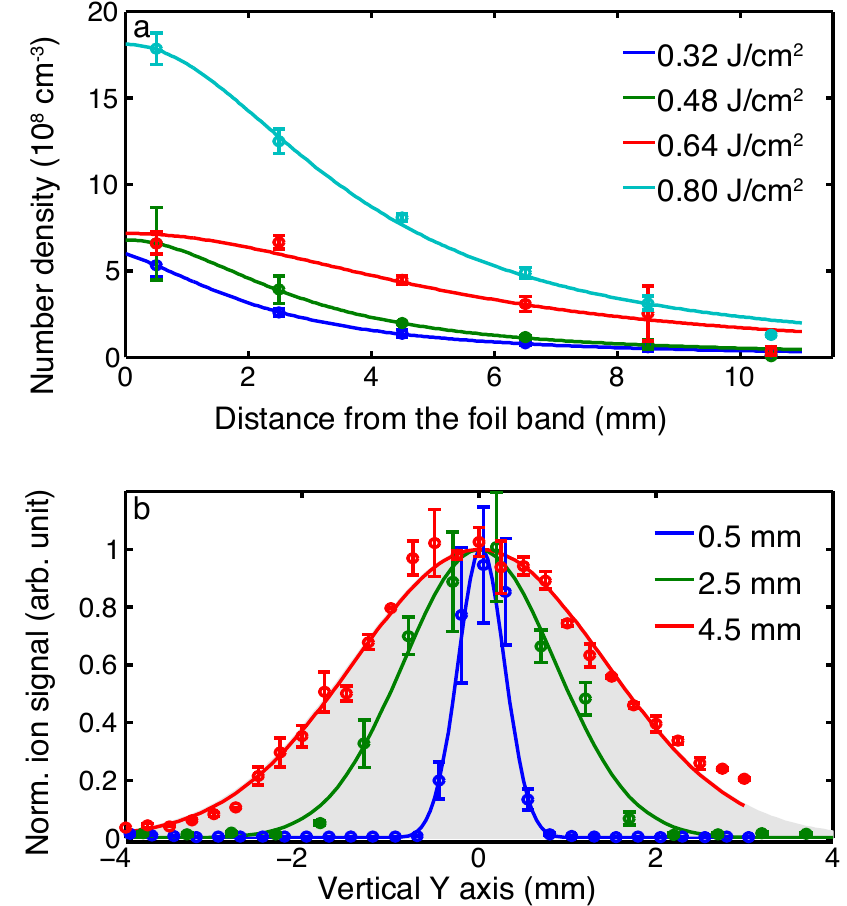}
   \caption{(a) Parent ion density as a function of distance from the foil band, showing
      inverse-square law behavior. (b) Transverse profile of the molecular plume at three distances
      from the foil band. Gray shading corresponds to the measured acceptance of the TOF
      spectrometer, such that the measurement at 4.5~mm does not represent the actual spatial extend
      of the plume, but the limits of the experimental acceptance. Solid lines correspond to
      Gaussian fits to the data.}
   \label{fig:spatial_profile}
\end{figure}

In the following we investigate the spatial extent, density, velocity, and translational temperature
of the ``plume'' of molecules desorbed from the foil band. We estimate absolute number densities
from ion counting measurements and the known interaction volume as defined by our ionization laser.
In \subautoref{fig:spatial_profile}{a} we show the measured number density of parent ions in the
center of the desorbed plume as a function of distance from the foil band. We note that the shown
densities are lower limits, since their calculation assumes an ionization efficiency of 1 for SFI
and considers the measured intact parent ions only, such that any fragmentation induced by the SFI
probe will reduce the derived density. The obtained densities exhibit approximately an
inverse-square-law behavior with distance from the foil, since the expansion along the laser
propagation direction is not reflected in the measurements due to the large Rayleigh length of the
ionization laser ($z_R\approx38$~mm). We note that the data point closest to the foil band for the
measurement at 0.64~J/cm$^2$ shows a significantly lower than expect density, which we can only
explain with a lower density of molecules attached on the desorption foil band for this measurement,
due to some instability during the aerosolization process.

We assess the spatial extent of the plume, \ie, the transverse profile, by translating the
ionization laser in height along the $y$-axis (\autoref{fig:setup}), across the plume of molecules.
This is shown in \subautoref{fig:spatial_profile}{b} for three distances between the foil band
surface and interaction point. The initial profile close to the foil band is very narrow, with a
FWHM of $\ordsim0.6$~mm after 0.5~mm of free flight. The plume then rapidly spreads out, reaching a
FWHM of around 2~mm after 2.5~mm propagation and within 4.5~mm of free flight the extent of the
plume exceeds the spatial acceptance of the TOF ion optics (indicated by the gray shading in
\subautoref{fig:spatial_profile}{b}), such that no accurate data can be measured at larger
separations. This rapid diffusion of the plume in space is consistent with the fast drop in density
observed as the distance between the foil band and the interaction point is increased,
\subautoref{fig:spatial_profile}{a}, and indicates rapid diffusion of the molecular plume in space
following desorption from a well-defined spot defined by the desorption laser profile.

\begin{figure}
   \centering
   \includegraphics{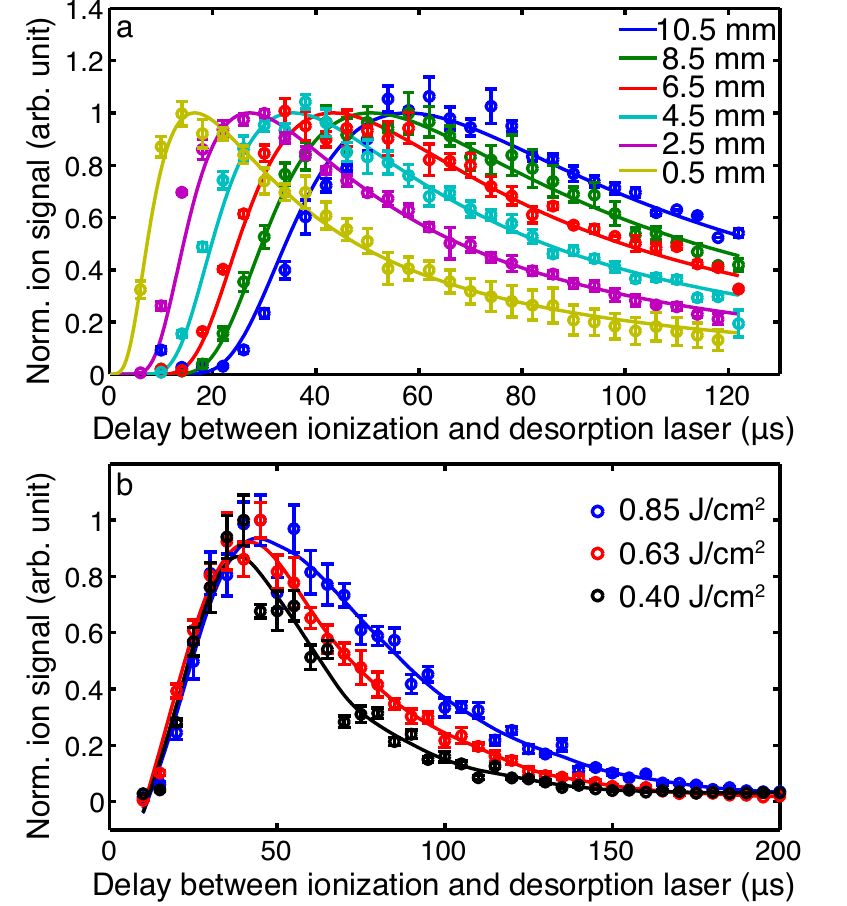}
   \caption{(a) Normalized temporal profiles of intact parent ions following desorption with
      0.8~J/cm$^2$, at different distances from the foil. Solid lines correspond to a fit with a
      Maxwell-Boltzmann distribution convoluted with the desorption time distribution. (b)
      Normalized temporal profiles of intact parent ions for different desorption laser intensities
      and otherwise identical settings, obtained at $z=6.5$~mm. While the most probable velocity is
      approximately constant, the larger desorption laser fluence leads to a much broader velocity
      distribution.}
    \label{fig:temporal_profiles}
\end{figure}
To investigate the longitudinal extend and velocity of the plume of desorbed molecules we measure
mass spectra as a function of delay between the desorption and ionization lasers, and at different
distances from the foil band. Results for the intact-parent-ion yield following desorption with a
fluence of 0.8~J/cm$^2$ are shown in \subautoref{fig:temporal_profiles}{a}. Similar data for other
desorption fluences are shown in the supplementary data. It is very clear that even when the
interaction point is very close to the foil band a broad temporal profile is observed, lasting
several tens of \us, much broader than the 8~ns duration of the desorption-laser pulse. At larger
distances from the foil band these distributions widen considerable more, demonstrating that during
free flight through the vacuum chamber the plume spreads out also in the longitudinal direction. We
identify two physical origins for the observed profiles and their temporal evolution; (i) the
desorption process itself that does not release molecules at one instant in time, but with a certain
temporal and kinetic energy distribution and (ii) the propagation of molecules in free flight with a
certain finite translational velocity distribution. Whereas (i) contains information about the
physical desorption mechanism from the foil, the translational velocity spread from (ii) corresponds
to the translational temperature in the moving frame of the molecules.

In order to accurately fit the measured data, one needs to convolute the initial desorption time
distribution from the foil band with the Maxwell-Boltzmann free-flight propagation. Since so far no
quantitative model is available to describe this desorption process accurately, we take the
experimental data measured closest to the foil band, \ie, 0.5 mm, as a measure of the initial
desorption time distribution and numerically convolute this with the Maxwell-Boltzmann model of the
free-flight propagation. Details of this convolution procedure and the Maxwell-Boltzmann model are
given in the supplementary information. We then perform a global fit of the data for all propagation
distances $l$ simultaneously using a common temperature $T$ and offset velocity $v_{0,z}$, while we
introduce only a single linear scaling parameter for the different data sets. The latter essentially
accounts for the drop in intensity along the probed center-line of the plume. The results of this
fit for a desorption laser fluence of 0.8~J/cm$^2$ are shown as solid lines in
\subautoref{fig:temporal_profiles}{a}, data for other fluences is provided in the supplementary
information. The obtained translational temperatures and forward velocities are summarized in
\autoref{tab:fitting}.
\begin{table}
   \centering
   \caption{Measured translational velocities and temperatures in the moving frame for different
      desorption laser intensities.}
   \label{tab:fitting}
   \begin{tabular}{*{3}{c}}
     \hline\hline
     Desorp. Fluence ~(J/cm$^2$) & ~~$T$~(K) ~~&  ~~$v_{0,z}$~(m/s)~~   \\
     \hline
     0.32 & $594$ & $233$   \\
     0.48 & $679$ & $234$   \\
     0.64 & $715 $ & $265$   \\
     0.80 & $758 $  & $224$  \\
     \hline\hline
   \end{tabular}
\end{table}

\begin{figure*}[t] 
   \centering
   \includegraphics{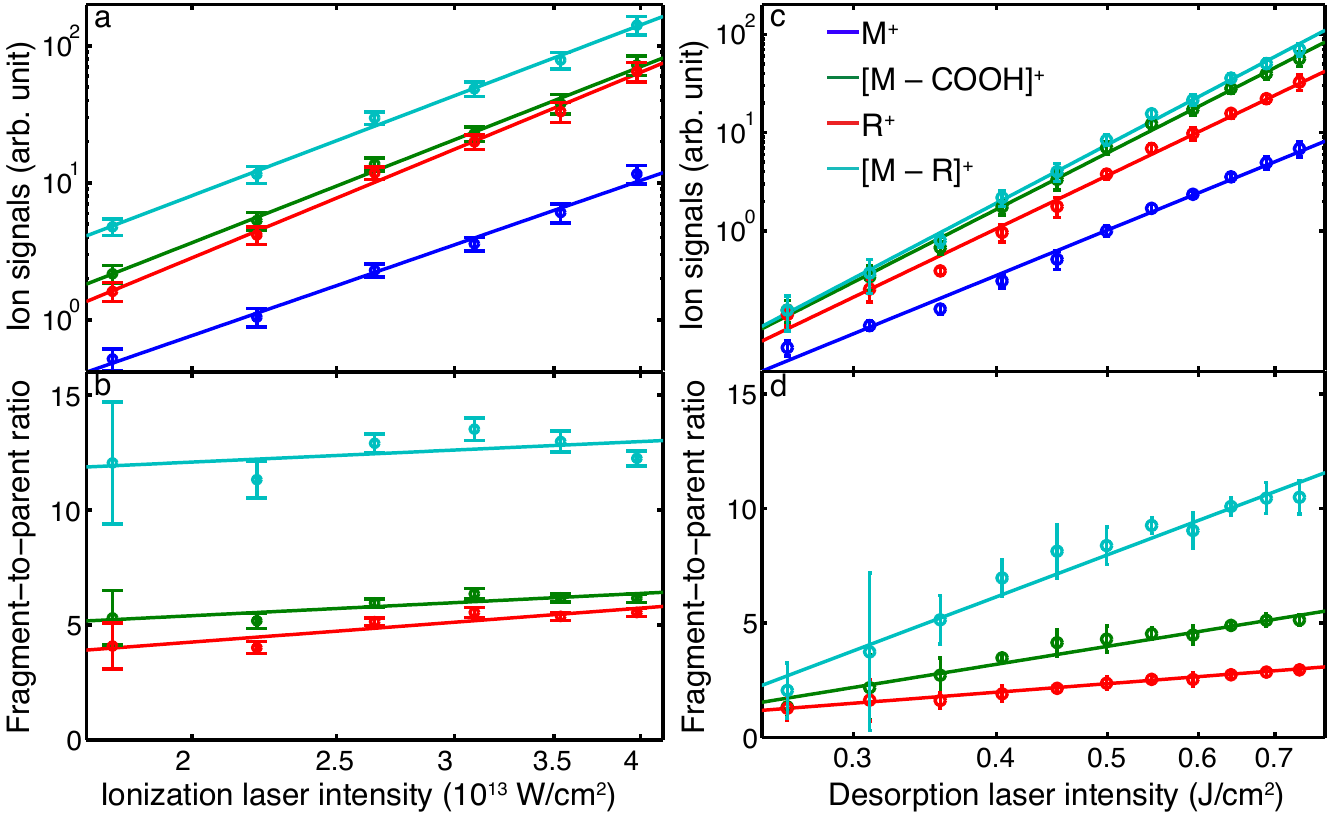}
   \caption{Ion-yield (a, c) and fragment-to-parent ratios (b, d) as a function of ionization laser
      intensity (a, b) and desorption laser intensity (c, d). Color coding for all graphs is given
      in panel c; see \autoref{fig:tof} for assignment of the mass peaks. Solid lines correspond to
      power-law fits.}
   \label{fig:intensity}
\end{figure*}

We observe a strong, nearly linear, dependence of the translational temperature of the molecular
plume on the fluence of the desorption laser. Even at the lowest fluence used a translational
temperature of nearly 600~K is obtained. In the current experimental setup using SFI we cannot
measure the internal (vibrational or rotational) temperature directly. However, given the large
density of states in systems such as phenylalanine, and the microsecond timescales of the desorption
process, we can assume a large degree of thermalization between the different degrees of freedom.
Thus the measured translational temperatures can be considered as a good indicator of the internal
temperature of desorbed molecules.

Unlike the temperature, the observed forward velocity appears to be approximately constant for the
different desorption laser fluences. The slightly elevated velocity for the measurement at
0.64~J/cm$^2$ could be due to instabilities in the sample preparation for this measurement, as
mentioned above. Similar observations of identical forward velocity have been previously
reported~\cite{Zinovev:AnalChem79:8232, Shea:AnalChem79:2688}. This invariability of the velocity
with desorption laser fluence suggests that this might be determined by material properties of the
substrate and the molecular sample.

\subautoref{fig:temporal_profiles}{b} shows the yield of intact parent ions as a function of
desorption laser-ionization laser delay for different desorption fluences. While the peaks of the
distribution overlap in time, the distribution is significantly broader for higher fluences. These
observations fully support our finding of a constant translational velocity, but increasing
translational temperature as the desorption laser fluence is increased (\emph{vide supra}).

\subsection{Molecular fragmentation}
\label{sec:fragmentation}
In how far the observed fragmentation is due to the desorption or the SFI process is hard to assess
from the mass spectra in \autoref{fig:tof} alone. In order to disentangle these contributions, we
collect mass spectra for different ionization and desorption laser intensities.

\subautoref{fig:intensity}{a} shows the ion yield for the PA parent and the three dominant
fragment ions as a function of ionization laser intensity, with all ion channels
showing a steep increase with increasing laser intensity. These data were fit with a power-law
dependence of the form $A\times{}x^n$. \subautoref{fig:intensity}{b} further shows the ratio of
fragment-to-parent ion abundances for the three dominant fragment ions, \ie, comparing
the relative abundances of the two respective channels. We observe only a very slight increase in
fragmentation as the laser intensity increases, in good agreement with previous studies suggesting
that SFI induced fragmentation is very sensitive to the employed pulse duration, but not the
intensity~\cite{Calvert:PCCP14:6289}.

\subautoref{fig:intensity}{c} shows the dependence of ion yields on the intensity of the desorption
laser and \subautoref{fig:intensity}{d} the corresponding fragment-to-parent ratios. The overall
measured ion abundances are again well described by a power-law fit and show a steep
increase for higher intensities, especially noticeable for fragment ions. This is confirmed by the
fragment-to-parent ratios, which also significantly increase with laser intensity, indicating
enhanced fragmentation. Thus, the desorption-laser interaction clearly induces fragmentation, either
directly during the desorption process or thereafter, but prior to ionization, \ie, as molecules
travel through the vacuum chamber toward the interaction point. To test the latter, we recorded mass
spectra at different distances behind the foil band, changing the laser-laser delay such that we
always probe the highest density part of the molecular plume, \ie, we follow the center of the plume
as it travels through the vacuum chamber.
\begin{figure}
   \centering
   \includegraphics[width=1\linewidth]{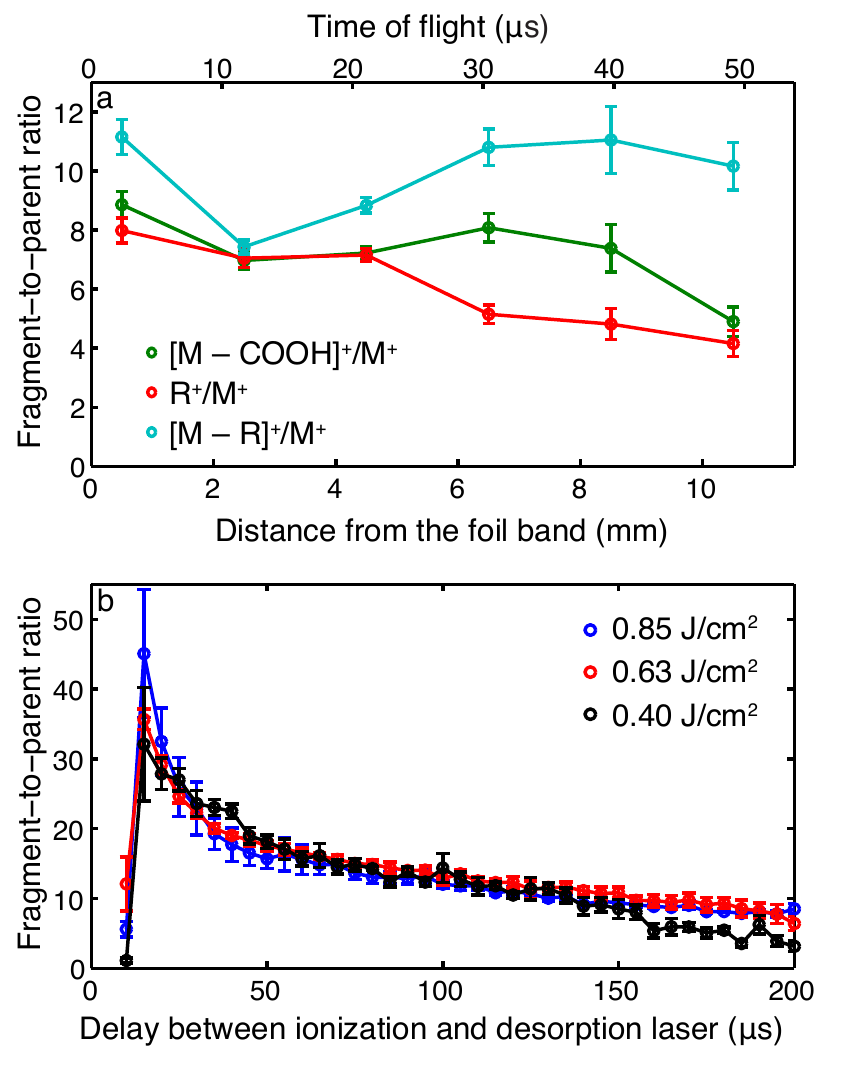}
   \caption{(a) Fragment-to-parent ratio recorded at the peak of the molecular plume density for
      different distances behind the foil. No significant increase in fragmentation is observed as
      the plume travels through the vacuum chamber. (b) Fragment-to-parent ratio throughout the
      molecular plume recorded 6.5~mm behind the foil. Molecules desorbed shortly after the arrival
      of the desorption laser show significantly higher fragmentation than molecules desorbed
      later.}
   \label{fig:fragmentation}
\end{figure}
This data is shown in \subautoref{fig:fragmentation}{a}, collected for distances of 0.5--10.5~mm
between the foil band and the interaction point, which corresponds to flight times of around
0--50~\us. Over this distance we observe no significant increase in fragmentation, indicating that
fragmentation occurs on much faster timescales, \ie, most likely during the desorption process
itself, either while molecules are still attached to the metal substrate or very shortly after
desorption into the gas-phase.

We now consider the distribution of fragments within a single plume coming from the foil band, \ie,
if the fragmentation changes depending on which part of the plume is observed. This is shown in
\subautoref{fig:fragmentation}{b}, where we plot the fragment-to-parent ratio for the
most abundant molecular fragment ion as a function of
desorption-laser-to-ionization-laser delay for a fixed distance from the foil band, \ie, 6.5~mm. We
observe an initial peak in the fragment-to-parent ratio at the onset of desorption, \ie, the
``front'' part of the molecular plume, which then decreases on a timescale of tens of microseconds.
These timescales are consistent with thermal processes, in particular we associate the observed
distribution with the rapid heating of the foil band by the nanosecond laser pulse, causing
increased fragmentation, followed by slow dissipation of the thermal energy, \ie, cooling down of
the front surface and, hence, reduced fragmentation. Further evidence that the fragmentation occurs
during the desorption process and that it is of a thermal nature comes from the comparison of the
fragment-to-parent ratios throughout the plume for different desorption laser fluences, also shown
in \subautoref{fig:fragmentation}{b}. These clearly show that the highest degree of fragmentation
occurs for the most intense desorption laser pulse. This is also consistent with the higher
translational temperatures derived for these conditions. Once the foil band cools down, \ie, at
longer desorption-laser-to-ionization-laser delays, the fragment-to-parent ratio approaches an
asymptotic value independent of initial desorption conditions.

\subsection{Nature of the desorption process}
Several possible mechanisms have been suggested in the literature for the underlying physical
processes occurring in the LIAD process~\cite{Lindner:IJMSIP103:203, Golovlev:APPL71:852,
   Zinovev:AnalChem79:8232, Goodfriend:APPA122:154}. It is important to note that the experimental
conditions for the different published LIAD-based molecule sources are very different;
pulsed~\cite{Calvert:PCCP14:6289, Zinovev:AnalChem79:8232} and
continuous~\cite{Calegari:Science346:336, Calegari:IEEESTQE21:1} desorption lasers are used and
sample preparation methods vary greatly, from the thick sample layer used here of
$\sim$500~nmol/cm$^2$~\cite{Calegari:IEEESTQE21:1, Borton:AnalChem85:5720}, to intermediate
thicknesses of tens of nmol/cm$^2$~\cite{Shea:AnalChem78:6133, Shea:AnalChem79:2688}, to
near-monolayer coverage in other studies~\cite{Zinovev:AnalChem79:8232}. As such, we do not aim to
provide a general model for the LIAD mechanisms, but seek to explain our observations and compare
these with previous studies where applicable.

One of the suggested desorption mechanisms, and indeed the origin of the term ``acoustic
desorption''~\cite{Lindner:IJMSIP103:203, Golovlev:APPL71:852}, is the direct momentum transfer from
a shock wave induced by the desorption laser in the foil band to the sample molecules. Our data
firmly rules out this mechanism for our molecule source. We observe a slow rise in molecular signal
on the order of $\ordsim10~\us$, see \autoref{fig:temporal_profiles}, which is not compatible with
molecules being ``shaken off'' by an impulse traveling through the foil, as this should lead to a
sharp sudden onset of signal as the impulse reaches the front surface, followed by an immediate drop
as the impulse is reflected on the surface. Additionally one might expect to observe a periodic
revival of signal as the impulses bounces back and forth within the metal foil. We observe no
evidence for this behavior. Furthermore, the travel time for a mechanical wave through a 10~\um
tantalum foil is approximately 2~ns~\cite{Rigg:JPCS500:032018}, significantly shorter than the delay
we observe between the desorption laser impacting on the foil and molecules being desorbed. A purely
acoustic desorption mechanism would, furthermore, not explain the observed increase in fragmentation
for increased desorption laser fluences. Similar observations have been made previously for a pulsed
LIAD setup, and the ``shake off'' mechanism similarly discredited~\cite{Zinovev:AnalChem79:8232}.

The other conceptually simple mechanism is a simple thermal one; the incident laser pulse heats up
the material from the backside and this thermal energy then diffuses to the front of the foil where
it heats up molecules and they eventually desorb. However, the observation that the velocity and,
therefore, the kinetic energy of desorbed molecules is independent of the incident desorption laser
power and thus surface temperature is not compatible with a purely-thermal desorption model.

The observation that the kinetic energy of desorbed molecules is independent of desorption laser
fluence indicates that this is determined by material properties of the foil substrate and/or the
molecular sample. This observation, along with the increase in translational temperature in the
moving frame, is consistent with a desorption model proposed by Zinovev
\etal~\cite{Zinovev:AnalChem79:8232}. They explain the LIAD process by an introduction of surface
stress between the substrate and the molecular sample -- located in isolated islands on the
substrate -- due to the acoustic and/or thermal wave created by the desorption laser. This surface
stress can lead to elastic deformation, decomposition, and cracking of sample islands on the foil
band and, eventually, to desorption of molecules. In this conceptual model the kinetic energy
transferred to a desorbing molecule is independent of the total incident laser power, and rather
depends on the intrinsic characteristics of a given sample island and substrate. A higher laser
fluence leads to the introduction of more surface stress and the formation of more cracks and
deformation sites, leading to an increase in molecular signal, but does not influence the amount of
kinetic energy per molecule. At the same time we note that due to thermal conductivity the higher
temperature of the substrate reached for higher desorption laser fluences will also heat up
deposited sample molecules due to thermal conduction, leading to internally hotter molecules,
increased fragmentation as well as higher translational temperatures.

While it is difficult to theoretically model the amount of energy transferred to each desorbed
molecule, Zinovev \etal provide a simple formula to estimate the energy per analyte molecule based
on material properties and thermal stress theory~\cite{Zinovev:AnalChem79:8232}. Based on this we
estimate 25--100~meV of energy per molecule for temperature differences of
$\Delta{}T=100$--200~K.~\footnote{We evaluated the energy release per molecule based on the known
   physical constants for anthracene~\cite{Bondi:JAP37:4643}, since data for PA was not available.
   The thermal expansion coefficient of the film is assumed to be
   $2.8\times10^{-4}~\mathrm{K}^{-1}$.} This is well within the range of the measured kinetic energy
per molecule which is, based on the average velocity observed, around 50~meV. Thus, our data is
fully supportive of the proposed surface stress model.

\section{Conclusion}
\label{sec:Conclusion}
We presented an advanced LIAD source for the preparation of gas-phase samples of labile molecules,
designed for the use at central-facility light sources such as free-electron lasers. It features a
prolonged continuous measurement time through automatic sample replenishment, as well as a fixed
interaction point. Uniform sample preparation on the long substrate was achieved using an aerosol
spraying method based on thin liquid jets. We have characterized the new source using phenylalanine
as a sample molecule and SFI as a universal probe method. We observe a significant fraction of
intact molecules being desorbed from the foil, with number densities around $2\times10^9$~cm$^{-3}$
close to the foil band. Due to fragmentation processes induced by the probe, this should be treated
as a lower limit. The spatial extend of the molecular plume rapidly spreads out from the point of
desorption, leading to a corresponding drop in density. The plume forward translational velocity and
temperature in the moving frame are derived by convoluting a Maxwell-Boltzmann velocity distribution
with the initial temporal profile near the foil band. The forward velocity, and hence kinetic
energy, of molecules desorbed from the foil does not depend on the desorption laser intensity. In
contrast to this, the translational temperature clearly increases with increasing desorption
intensity. We investigated the fragmentation processes and observe increased fragmentation at higher
desorption laser intensity, consistent with the translational temperature behavior. Furthermore, we
show that the amount of fragmentation depends on the time of desorption from the foil: shortly after
the laser pulse molecules are observed to be hottest, and subsequently they cool down on thermal
timescales (10s of \us) as the substrate itself cools down. These observations are fully supported
by the previously proposed surface-stress model of the LIAD process.

Our characterization measurements show that our new source produces a stable high-density signal of
intact molecules in the gas-phase. With automatic sample replenishment it provides very long
continuous measurement times. The produced molecular plume is well suited for further gas-phase
experiments and manipulation, and work is currently underway towards integrating this source into a
buffer-gas-cooling setup for the production of cold molecules~\cite{Hutzler:CR112:4803}, which can
then be further manipulated using electric fields~\cite{Chang:IRPC34:557}. One could also envision
to make use of this desorption technique for the entrainment of molecules into supersonic beams,
similar to matrix-assisted laser desorption approaches~\cite{Teschmit:JCP147:144204}.

  \section{Supporting Information}
  The supporting information contains details regarding the
  \begin{itemize}
     \itemsep=0pt
  \item Experimental setup
  \item Sample preparation and deposition
  \item Derivation of Maxwell-Boltzmann velocity distributions
  \item Temporal profiles at different desorption intensities
  \end{itemize}

\begin{acknowledgments}
   In addition to DESY, this work has been supported by the European Research Council under the
   European Union's Seventh Framework Programme (FP7/2007-2013) through the Consolidator Grant
   COMOTION (ERC-614507-Küpper), by the excellence cluster ``The Hamburg Center for Ultrafast
   Imaging -- Structure, Dynamics and Control of Matter at the Atomic Scale'' of the Deutsche
   Forschungsgemeinschaft (CUI, DFG-EXC1074), and by the Helmholtz Gemeinschaft through the
   ``Impuls- und Vernetzungsfond''. Z.\,H.\ gratefully acknowledges a scholarship of the
   Joachim-Herz-Stiftung and support by the PIER Helmholtz Graduate School.
\end{acknowledgments}

\vspace*{1em}

\bibliographystyle{achemso} 
\bibliography{string, cmi}


\end{document}